\documentclass[preprint,tightenlines,showpacs,showkeys,floatfix,nofootinbib,superscriptaddress,fleqn]{revtex4} 
\usepackage{bm}
\usepackage{amsmath}
\usepackage{amsfonts}
\usepackage{amssymb}
\usepackage{amscd}
\usepackage{amsthm}
\usepackage{epsf}
\newcommand{\ds}{\displaystyle}
\newcommand{\dsf}{\ds\frac}
\newcommand{\Tr}{\mbox{Tr\,}}
\newcommand{\re}[1]{(\ref{#1})}
\begin{document}
\sloppy
\preprint{PNU-NTG-11/2004}
\title {Is there a crystalline state of nuclear matter?}
\author{U.T. Yakhshiev}
\email{yakhshiev@pusan.ac.kr}
\affiliation{Department of Physics and Nuclear Physics \& Radiation
Technology Institute (NuRI),\\ Pusan National University, 609-735
Busan, Republic of Korea}
\affiliation{Theoretical Physics Department \& Institute of Applied
  Physics\\ National University of Uzbekistan, Tashkent-174,
  Uzbekistan} 
\author{M.M. Musakhanov}
\email{musakhanov@pusan.ac.kr}
\affiliation{Department of Physics and Nuclear Physics \& Radiation
Technology Institute (NuRI),\\ Pusan National University, 609-735
Busan, Republic of Korea}
\affiliation{Theoretical Physics Department \& Institute of Applied
  Physics\\ National University of Uzbekistan, Tashkent-174,
  Uzbekistan} 
\author{H.-Ch. Kim}
\email{hchkim@pusan.ac.kr}
\affiliation{Department of Physics and Nuclear Physics \& Radiation
Technology Institute (NuRI),\\ Pusan National University, 609-735
Busan, Republic of Korea}

\date{March 2005}

\begin{abstract}
A possibility of the crystalline state of nuclear matter is discussed in a
medium--modified Skyrme model.  The interaction energy per nucleon in
nuclear matter is evaluated by taking into account the medium
influence on single nucleon--skyrmion properties and the tensor part
of the nucleon--nucleon potential, and by using a variational method
of Hartree--Fock type including zero--point quantum fluctuations.  It
is shown that in this approach the ground state of nuclear matter
has no crystalline structure due to quantum fluctuations as well as
medium modifications of hadron properties. 
\end{abstract}

\maketitle

{\bf 1.} After a pioneering work by Kutchera {\em et al.}~\cite{Kutschera84} 
related to the dense--matter properties
in the Skyrme model, Klebanov~\cite{Klebanov85} discussed a
possible formation of the Skyrmionic matter with the simple cubic
crystalline structure due to tensor forces between the nucleon
of a lattice--point and the nearest six nucleons of the unit cell.
There has been a great amount of work on energetically favorable crystalline
structures~\cite{Glendenning86,Wust87,Jackson88} and   
parallel studies of the Skyrme model on the hypersphere~\cite{Manton86,Jackson,JaWiMa}.
Former studies are still under discussion~\cite{Lee} and
have concentrated on investigating the classical field configuration,
assuming face centered cubic (FCC) structure.  Obviously, one should
quantize the model to investigate the real nuclear  
matter, but it is not at all easy to do it as explained in
Ref.~\cite{Park:2002ie}. 

On the other hand, quantum fluctuations around a classical ground
state can play an important role in leading to nuclear matter which is
different in its behavior from the classical Skyrmion matter.  A
standard variational method of Hartree--Fock type used in
Ref.~\cite{Diakonov88} provides a theoretical framework to make it
tractable to deal with nuclear matter with quantum fluctuations.
Taking into account only the tensor part of the nucleon--nucleon (NN)
interaction based on the one-pion-exchange potential (OPEP) and
assuming the FCC structure, Diakonov and Mirlin~\cite{Diakonov88} 
showed that at densities which are several times higher that that of ordinary
nuclear matter a condensed state of nuclear matter may be still
energetically favorable.  Alternatively, using time-dependent
numerical calculations on a quantized simple cubic lattice,
Walhout~\cite{Walhout88} evaluated the zero-point kinetic  
energy of nucleons and found it to unbind the classical crystalline
ground state.  Consequently, one can conclude that due to the
increasing number of neighboring nucleons the FCC structure appears in 
a deeper bound state.  

However, above-mentioned studies did not consider possible
medium modifications of single nucleon and pion properties and the 
corresponding medium--modified NN 
interaction~\cite{Durso:1992hu,prc001,npa001,npa002}.
Refs.~\cite{prc001,npa001,npa002} considered the nucleon as a 
Skyrmion and then studied the influence of baryonic matter on its
properties phenomenologically via medium modifications  
of pion fields.  The results of these studies were in a qualitative
agreement with experimental indications, e.g. the swelling of the
nucleon in nuclear medium and the decrease of its mass.  It was also
shown that due to the influence of surrounding nuclear environment the 
tensor part of the NN potential is decreased~\cite{prc001}.  The 
changes in hadron static properties and form factors of
meson--nucleon vertices also lead to the suppression
of the one-boson-exchange potential~\cite{npa001}.  In this context,
it is natural to ask and answer the question: How do these changes play   
a role in describing nuclear matter?

Since it is of great difficulty to quantize the Skyrmionic matter
itself~\cite{Park:2002ie}, we follow the phenomenological and
variational method put forward by Ref.~\cite{Diakonov88} in order to
investigate how the medium modifications of the NN potential and of
hadronic properties change the ground-state energy of nuclear matter
and reestimate the role of zero-point quantum fluctuations around the 
minimum.

\vspace{1cm}
{\bf 2.} In order to calculate the medium--modifications of single
hadron properties, we start with the effective chiral Lagrangian for the
medium--modified Skyrmion~\cite{prc001}.  Using the phenomenological
values $b_0=b_0^{phen}=-0.024m_\pi^{-1}$, $c_0=c_0^{phen}=0.15m_\pi^{-3}$ of
the effective pion--nucleon scattering lengths~\cite{Ericson}, one
gets the density dependence of hadron properties as listed in
Table~\ref{medpar}.\footnote{We refer the reader to the
works~\cite{prc001,npa001}, where the medium modifications of
hadronic properties have been investigated and inclusion of medium parameters 
($b_0$, $c_0$ and $g'$) were discussed.} 
\begin{table}[htb]
\begin{tabular}{l|c|ccc|ccc}\hline\hline
\quad$\rho/\rho_0$ &  0 &\multicolumn{3}{|c|}{0.5} & \multicolumn{3}{|c}{1.0}\\
\hline
\quad$g'$ & - & 0.33 & 0.6 & 1 & 0.33 & 0.6 & 1 \\
\hline
$g_{\pi NN}^*$ & 12.49 & 9.48 & 9.76 & 10.08 & 6.83 & 7.75 & 8.66 \\
$M_N^*$[MeV]   & 868 & 743 & 756 & 770 &  635 & 675 & 715 \\
$m_\pi^*$[MeV] & 140 & 146 & 146 & 146 & 152 & 152 & 152 \\
$\Lambda^*$[MeV] & 528 & 484 & 489 & 494 & 448 & 462 & 477 \\
$M_{\Delta N}^*$[MeV] & 243 & 211 & 214 & 218 & 186 & 196 & 206\\
\hline\hline
\end{tabular}
\caption{\label{medpar}
Comparison of hadronic properties in free space with those in
nuclear matter with density $\rho$.  In the second column
the corresponding values in free space are listed, $g'$ stands for the
correlation parameter, $\rho_0$=0.17~fm$^{-3}$ is the density of
normal nuclear matter.  $M_{\Delta N}^*\equiv M_{\Delta}^*-M_N^*$ 
is the $\Delta$-N mass splitting.}
\end{table}
In the present work, the asterisk in notation indicates
medium--modified quantities.  The results are presented for three
different values of the Lorentz--Lorenz parameter:
the first and the second ones, i.e. $g'=0.33$ and $g'=0.6$, 
correspond to the classical and phenomenological
values~\cite{Ericson}, respectively, and 
the third one is simply taken as $g'=1$.  One can see
that hadronic properties as well as the pion--nucleon
coupling constant are changed due to the presence of nuclear
medium.  Consequently, the nuclear--matter properties, being calculated
by using the NN potential between the nearest neighbors in nuclear matter,
are expected to be modified accordingly.  

In order to study a phase structure of nuclear matter, we follow
closely a variational procedure described in Ref.~\cite{Diakonov88}.
We define the position vector of an FCC lattice point $n$ by
$\bm r_{0n}$ and the optimum orientation (the relative
orientation matrix in internal space) of the nucleon
at this lattice point by $A_{0n}$.  The general orientation matrix $A$ 
is related to an arbitrary position vector in isospin space.  The
trial wave function of the crystal must be chosen as
an antisymmetrized product of single-lattice-point
wave functions $\psi(\bm r-\bm r_{0n}, A^+A_{0n})$
localized near the lattice sites and the corresponding optimum
orientations.

The overlap integrals of the non-nearest neighbors being neglected,
the energy functional (interaction energy per nucleon) 
can be written as a following Hartree--Fock form:
\begin{equation}
\begin{array}{l}
E^*=\ds\int d^3r~dA~\psi^*(\bm r-\bm r_0,A^+A_0)
\left(-\dsf{1}{2M_N^*}\dsf{\partial^2}{\partial \bm r^{\,2}}
+\dsf{\bm T_A^{\,2}}{2I^*}\right)\psi(\bm r-\bm r_0,A^+A_0)\\
+\dsf{1}{2}\sum_{n=1}^{12}\int d^3r~d^3r'~dA~dA'
\psi^*(\bm r-\bm r_0,A^+A_0)\psi^*(\bm r\,'-\bm r_{0n},A^{\prime+}A_{0n})
V^*(\bm r-\bm r\,',A^+A')\\
\times
\left[\psi(\bm r-\bm r_0,A^+A_0)\psi(\bm r\,'-\bm r_{0n},A^{\prime+}A_{0n})
-\psi(\bm r-\bm r_{0n},A^+A_{0n})\psi(\bm r\,'-\bm r_0,A^{\prime+}A_{0})
\right]\,,
\end{array}
\label{H-F-f}
\end{equation}
where $I^*$ denotes the moment of inertia of the Skyrmion in medium.
This functional must be minimized with respect to the
single-lattice-point wave function $\psi$ for a given distance $R=|\bm
r_{0n}-\bm r_0|$ between the nearest neighbors.  It is clear that in
general the quantities, $M_N^*,I^*,V^*$, are $R$-dependent, while the 
$R$ is related directly to the nuclear-matter density.

We use the single-lattice-point trial function as follows:
\begin{eqnarray}
\psi(\bm r-\bm r_{0},A^+A_{0})=\varphi(\bm r-\bm r_{0})
\dsf{D_{kk}^{1/2}(A^+A_0)+\alpha D_{kk}^{3/2}(A^+A_0)}
{\sqrt{1+\alpha^2}}\,,
\label{psibegin}\\
\varphi(\bm r\,)=
\left({\varkappa^2}/{\pi}\right)^{3/4}
\left(1+3\eta+{15\eta^2}/{4}\right)^{-1/2}
\exp\left\{-{\varkappa^2 r^2}/{2}\right\}(1+\eta\varkappa^2r^2)\,,
\label{varphi1}
\end{eqnarray}
where $\varkappa$ measures the spatial extent of the wave function,
$\alpha$ denotes the admixture of the $\Delta$-resonance,
$D^T_{ab}(A)$ is an SU(2) Wigner $D$ function, and $\eta$   
stands for an additional variational parameter.  Putting $\eta=0$,
one can get the corresponding trial function similar to
Ref.~\cite{Diakonov88}. 

In general, NN potential can be presented in the following
form:
\begin{equation}
V^*(\bm R,A_1^+A_2)
=V_1^*(R)+\dsf{1}{2}\,\Tr(A_1^+A_2\sigma_iA_2^+A_1\sigma_j) 
[(3\hat R_i\hat R_j-\delta_{ij})V_2^*(R)+\delta_{ij}V_3^*(R)]\,,
\label{NNpotential}
\end{equation}
where $\hat R$ is the corresponding unit vector
along the line joining the centers of two nucleons
and $A_1^+A_2$ denotes their relative orientation matrix in internal
space.  Utilizing Eqs.~\re{psibegin}--\re{varphi1} of 
the lattice-point trial wave function and the NN potential
given in the form of Eq.~\re{NNpotential}, one can arrive at
the final expression for the energy functional (Eq.~\re{H-F-f}): 
\begin{eqnarray}
E^*&=&
\dsf{\alpha^2}{1+\alpha^2}M_{\Delta N}^*
+\left(1+3\eta+\dsf{15\eta^2}{4}\right)^{-1}
\left(1+\eta+\dsf{11}{4}\eta^2\right)
\dsf{3\varkappa^2}{4M_N^*}+E_H^{*}+E_F^{*}\,,
\nonumber\\
E_H^{*}&=&
\dsf{3}{\pi^2}\left(1+3\eta+\dsf{15\eta^2}{4}\right)^{-2}
\int\limits_{0}^{\infty}dq~q^2
\exp\left\{-\dsf{q^2}{2\varkappa^2}\right\}
\nonumber\\
&&\times\,\, a^2\left[V_1^*(q)+\dsf{(1+\alpha)^2}{9(1+\alpha^2)^2}
\left(\dsf{V_2^*(q)}{q^2}
\dsf{\partial^2}{\partial R^2}-V_3^*(q)\right)\right]\dsf{\sin qR}{qR}
\,,\nonumber\\
E_F^{*}&=&-\dsf{3}{\pi^2}
\left(1+3\eta+\dsf{15\eta^2}{4}\right)^{-2}
\left[\dsf{4(1-\alpha-\alpha^2)}{3(1+\alpha^2)}\right]^2\int\limits_0^\infty
dq~q^2
\exp\left\{-\dsf{\varkappa^2R^2}{2}-\dsf{q^2}{2\varkappa^2}\right\}
\nonumber\\
&&\times
\left[\left(\dsf{b^2}{3}+\dsf{4bc}{15}+\dsf{3c^2}{35}\right)V_2^*(q)+
\left(b^2+\dsf{2bc}{3}+\dsf{c^2}{5}\right)V_3^*(q)\right]\,,
\label{FinEnFunc}
\end{eqnarray}
where the variables $a$, $b$, and $c$ are defined as
\begin{eqnarray}
a&\equiv&\left[1+\left(6-{q^2}/{\varkappa^2}\right){\eta}/{2}+
\left(60-20{q^2}/{\varkappa^2}+{q^4}/{\varkappa^4}\right)
{\eta^2}/{16}\right]\,,\nonumber\\
b&\equiv&a+R^2\varkappa^2[{\eta}/{2}+(4+R^2\varkappa^2-2q^2/\varkappa^2)
{\eta^2}/{16}]\,,\nonumber\\
c&\equiv& 4q^2R^2{\eta^2}/{16}\,.
\end{eqnarray}

To compute the ground-state energy in nuclear matter, we keep only a
tensor part of the potential in Eq.\re{NNpotential}, putting 
$V_1^*=V_3^*=0$.  This tensor potential arises from one-pion exchange,
which is expressed in momentum space as follows:
\begin{equation}
V_2^*(q)=-\left(\dsf{3g_{\pi NN}^*}{2M_N^*}\right)^2\dsf{\bm
q^2}{\bm q^2+m_\pi^{*2}}\, 
\dsf{\Lambda^{*2}}{\bm q^2+\Lambda^{*2}}\,.
\label{OPEP}
\end{equation}

The effective values of the potential parameters
and their density dependence can be calculated by using
the medium--modified Skyrme Lagrangian given in Ref.~\cite{prc001}\footnote{ 
The medium--modified values of the OPEP parameters at some densities
are presented in Table~\ref{medpar}, where $g_{\pi NN}^*$ and $\Lambda^*$
are  extracted from the
medium--modified $\pi NN$ form factor (see Ref.~\cite{npa001} for 
more details).}.
We present two different sets of OPEP parameters in free space:
OPEP~I ($g_{\pi NN}$=14, $M_N$=940~MeV, $m_\pi$=140~MeV,
$\Lambda$=500~MeV, $M_{\Delta N}$=300~MeV) used in 
Ref.~\cite{Diakonov88} and OPEP~II which coincides with the second
($\rho=0$) column of Table~\ref{medpar}. We emphasize that our aim is
not to describe quantitatively the NN potential in free space but to
study qualitative changes in interaction energy due to the medium
modifications.  

\begin{figure}[hbt]
\vskip 0.5cm
   \epsfysize=6.5cm\centerline{\epsffile{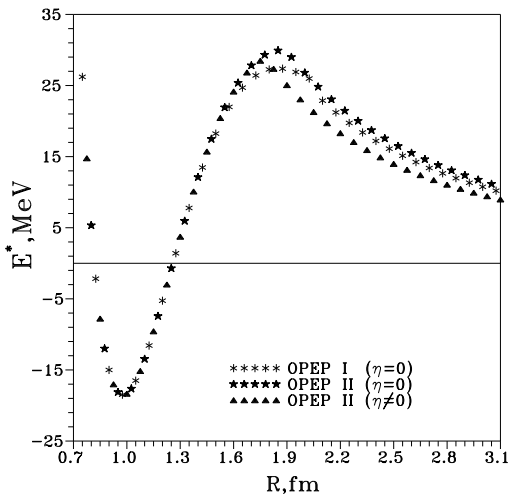}}
\caption{
\label{opep-grf}\small
The minimization of the interaction energy $E^*$ (Eq.~\re{FinEnFunc})
by using two sets of the OPEP parameters.
The OPEP~I ($\eta=0$) corresponds to the results
of Ref.~\cite{Diakonov88}.  Those with the OPEP II are drawn for
two different values of the additional variational
parameter, which is defined in Eq.~\re{varphi1}, i.e.
$\eta=0$ and $\eta\ne 0$, respectively.} 
\end{figure}

These two parameter sets show almost the same
dependence on $R$ for the interaction energy per nucleon as illustrated in
Fig.~\ref{opep-grf}.  The curve with asterisks which corresponds to
the OPEP~I set with $\eta=0$ is equivalent with that of
Ref.~\cite{Diakonov88}.  The curve with stars corresponding to the
OPEP~II with $\eta=0$ draws the result of the present calculation.
Though we are not able to distinguish the results with the OPEP~II
from those with the OPEP~I in shorter ranges, they start to get
deviated each other when $R$ grows:
The $\pi NN$ coupling constant in the
OPEP~I set is larger than that in the OPEP~II, so that the interaction
energy becomes more attractive.  Note that the solutions with the OPEP~II 
($\eta\ne 0$) are stable with respect to the change of the
lattice--point nucleon trial functions.  The additional parameter
$\eta$ plays an important role only at larger distances, i.e. $R\ge
1.6$~fm, where the energy is getting lower.    

In order to see how the ground-state energy of nuclear matter is
changed with the medium modifications of the potential
parameters taken into account, we begin with the OPEP~II for both
$\eta=0$ and $\eta\ne 0$ cases so that we can keep our investigation
in a self-consistent manner.  The density of the medium in the FCC
structure is defined as $\rho=\sqrt{2}/R^3$~\cite{Diakonov88}.  The
normal nuclear--matter density $\rho=0.17$~fm$^{-3}$ corresponds to a 
distance between the nearest neighbors: $R\approx 2.03$~fm.  The
medium modifications of the single hadron properties are expressed via  
three phenomenological quantities such as $b_0$, $c_0$, $g'$ for a
given density of nuclear matter~\cite{prc001}.  We now treat the
phenomenological effective $S$- and $P$-wave pion-nucleon scattering lengths $b_0$,
$c_0$, and the correlation parameter $g'$ in deriving the interaction
energy per nucleon, respectively.     
\begin{figure}[hbt]
\vskip 0.5cm
   \epsfysize=6.5cm\centerline{\epsffile{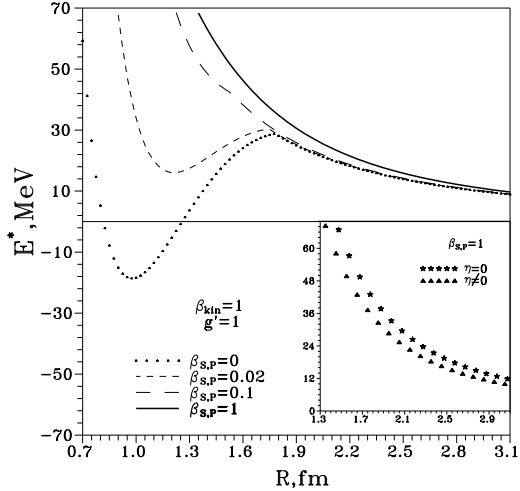}}
\caption{\label{b0c0}\small
The effect of the medium modifications of hadron properties
on the interaction energy per nucleon.  The dotted curve represents the
case, where the medium modifications of hadron properties are not
taken into account ($\beta_{S,P}=0$), while the short-dashed one
depicts the case when we consider the empirical parameters $b_0$ and
$c_0$ by $2\%$, i.e. $\beta_{S,P}=0.02$.  The long-dashed curve draws
the result with $\beta_{S,P}=0.1$, whereas the solid one takes into
account $\beta_{S,P}=1$, i.e. the full effect of the medium 
modifications.  In these curves, we always consider $\eta\ne 0$.  In
the small panel, two different curves are shown in the case of
$\beta_{S,P}=1$, i.e. $\eta=0$ and $\eta\ne 0$, 
respectively.}
\end{figure}
While $g'$ is fixed like $g'=1$, we can examine how the interaction
energy per nucleon depends on the parameters $b_0$ and 
$c_0$.  Introducing the controlling variables $\beta_S$ and $\beta_P$,  
we can rewrite $b_0$ and $c_0$ as follows:
\begin{equation}
b_0=b^{phen}_0\beta_S\,,\qquad c_0=c^{phen}_0\beta_P\,.
\label{bc-control}
\end{equation}
They allow us to include the medium effects in a controlled manner: 
$\beta_{S,P}$ are varied in the range of  $0\le \beta_{S,P}\le 1$.
Putting $\beta_S=\beta_P=0$, we restore them in free space.    

Similarly, we introduce the overall kinetic factor $\beta_{kin}\le 1$
in the energy functional~\re{FinEnFunc} to show the importance of
zero--point kinetic and iso-rotational ($\Delta$ admixture)
fluctuations:  
\begin{equation}
E^*=
\left[\dsf{\alpha^2}{1+\alpha^2}M_{\Delta N}^*+
\left(1+3\eta+\dsf{15\eta^2}{4}\right)^{-1}
\left(1+\eta+\dsf{11}{4}\eta^2\right)\dsf{3\varkappa^2}{4M_N^*}
\right]\beta_{kin}+E_H^{*}+E_F^{*}\,.
\label{toten}
\end{equation}

\vspace{1cm}
{\bf 3.} The effect of the medium parameters $b_0$, $c_0$ on the interaction
energy per nucleon (Eqs.~\re{FinEnFunc}, \re{toten}) is depicted in 
Fig.~\ref{b0c0}.  If the medium modifications are ignored, the quantum
fluctuations are not enough to loose a close-packed FCC structure.  Thus, 
nuclear matter can be formed as a FCC crystal when the
density is several times higher than that of normal nuclear
matter~\cite{Diakonov88}, as shown in the dotted curve
($\beta_{S,P}=0$) in Fig.~\ref{b0c0}.  With $\beta_{S,P}$ switched on,
one can see that the corresponding modifications of the NN tensor
potential essentially changes this picture of nuclear matter.  It
implies that including a small amount of the medium modifications
breaks the condensed (or solid) FCC structure, as presented in
short and long dashed curves in Fig.~\ref{b0c0}.  It is found that
$\beta_{S,P}=0.1$ breaks already the FCC structure.  Consequently, the
full consideration of the medium modifications, i.e. $\beta_{S, P}=1$ 
does not allow nuclear matter to be found in the form of a
FCC crystal.  This result is drawn in the solid curve in
Fig.~\ref{b0c0} and stable under the change of $\eta$, which is shown  
in the small panel of Fig.~\ref{b0c0}.    

\begin{figure}[hbt]
\vskip 0.5cm
   \epsfysize=6.5cm\centerline{\epsffile{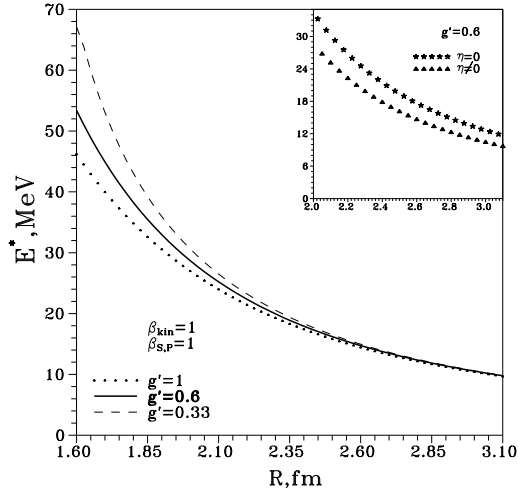}}
\caption{
\label{g'}\small
The effect of the Lorentz--Lorenz parameter $g'$ on
the interaction energy per nucleon.
Dotted, solid, and dashed curves correspond to the results
for $g'=1$, $g'=0.6$ (phenomenological value), and $g'=0.33$ (classical
value), respectively.  The medium modifications of the OPEP parameters
are fully taken into account, i.e. $\beta_{S,P}=1$.  The main panel
represents the case with  $\eta\ne 0$.  In the small panel, the
results are presented for both $\eta = 0$ and $\eta\ne 0$ cases with
$g'=0.6$.}
\end{figure}

The effect of the Lorentz--Lorenz parameter on the interaction
energy per nucleon is drawn in Fig.~\ref{g'}.
We present here the results with 
quantum fluctuations and medium modification effects fully taken into
account ($\beta_{kin}=\beta_{S,P}=1$).  As shown in Fig.~\ref{g'}, the
interaction energy per nucleon depends on the Lorentz--Lorenz
parameter rather weakly.  Moreover, 
when $g'$ decreases in such a way that it is taken to be 
the empirical ($g'=0.6$) or classic ($g'=0.33$), it is even more
difficult to find the crystalline structure of the ground state.  All
curves in the main panel correspond to the full variational 
function, including parameter $\eta$.  The
small panel of Fig.~\ref{g'} demonstrates the role 
of the variational parameter $\eta$, where $g'=0.6$ is chosen.  

The present results imply that the medium--modifications dramatically
changes the picture of nuclear matter due to the following reasons:
Firstly, the modifications of single-nucleon properties
make the contribution of the quantum fluctuations increased, since
the nucleon mass is decreased in the kinetic term (see
eq.~\re{toten}).  Secondly, the tensor
part of the NN potential is suppressed in nuclear matter.  As a
result, its negative contribution to the total interaction energy is
getting smaller.

\begin{figure}[hbt]
\vskip 0.5cm
   \epsfysize=6.5cm\centerline{\epsffile{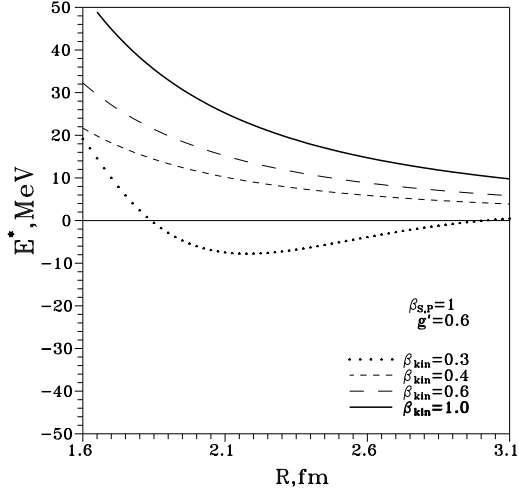}}
\caption{\label{QF}\small
The effect of the parameter $\beta_{kin}$ on the interaction
energy per nucleon~\re{toten}.  The medium modifications of 
hadron properties are fully taken into account ($\beta_{S,P}=1$) and 
the Lorentz--Lorenz parameter is fixed to be $g'=0.6$.}
\end{figure}

It is interesting to study whether the medium modifications
play a similar role, when the quantum fluctuations are suppressed.  
In order to see the effect of the medium modifications of hadron
properties, we artificially vary the parameter $\beta_{kin}$ to
suppress the quantum fluctuations.  The results are drawn
in Fig.~\ref{QF}.  
The suppression of the quantum fluctuations leads obviously to
lowering the interaction energy per nucleon.  This effect is shown in 
the long-dashed and short-dashed curves in Fig.~\ref{QF}.  When the
quantum fluctuations become weak enough ($\beta_{kin}\approx 0.3$),
the ground-state energy of the system makes it possible for nuclear
matter to be in the solid or condensed FCC
configuration, as discussed already in 
Refs.~\cite{Glendenning86,Wust87,Jackson88,Diakonov88}.  
This is shown by the dotted curve in Fig.~\ref{QF}.  

In the present work, we investigated the structure of nuclear matter
in the framework of the medium-modified Skyrme model.  We found that
the medium modifications of the hadron properties and    
NN tensor potential are crucial to see whether the crystalline
structure of nuclear matter exist or not.  It turned out that the
contribution of the medium modifications to the interaction energy per
nucleon  breaks a possible crystalline structure of the ground state
of nuclear matter.  We found that the present results are stable under
the changes of numbers of variational parameters in the lattice-point
nucleon trial functions.    

However, one has to note that there might be a possibility that a
condensed state at the normal nuclear--matter density could exist
within the framework of the medium-modified Skyrme model according to
the following reasons: Firstly, because of the contribution of the medium
modifications to the interaction energy,  a {\em desirable} minimum of 
nuclear matter seems to appear around the normal nuclear--matter
density if quantum fluctuations are suppressed.  The dotted curve in
Fig.~\ref{QF} indicates this possibility.  However, we want to
emphasize that without the  
medium modifications one should not conclude the existence of the
crystalline structure of nuclear matter.  Secondly,
since it is well known that the binding of nuclear
matter results from a strong cancellation between an attractive
potential term and a repulsive kinetic one, the present conclusion may
be changed if one can include the central attractive NN potential,
which is in the present work neglected.  For example, it is asserted
in Ref.~\cite{Serot86} that the scalar-isoscalar
degrees of freedom are important.  One can include the contribution of
the scalar-isoscalar channel in the Skyrme model in a similar way as
done for the pseudoscalar-isovector channel.  For example,
Refs.~\cite{Durso:1992hu,npa001,epja003} showed that the attraction of
the central NN potential is modified in nuclear matter in such a way
that it is suppressed as the density increases.  It is clear that due to
these modifications the whole attractive potential containing the
central and tensor parts will be lessened.  However, if 
the scalar degrees of freedom are added in the medium-modified Skyrme
model, the nucleon effective mass is less changed~\cite{npa001}.
Consequently, the kinetic part on the interaction energy should be 
suppressed (quantum fluctuations will be suppressed). Therefore, it
will be of great importance to understand the interplay between these
terms in the interaction energy within the framework of the
medium-modified Skyrme model, the scalar-isoscalar channel being
considered.  

\section*{Acknowledgments}
The present work is supported by the Korea Research Foundation Grant
(KRF-2003-070-C00015).  The work of UTY was financially supported by Pusan
National University in program Post-Doc. 2004.  MMM acknowledges the
support of the Brain Pool program 2004 which makes it possible for him
to visit Pusan National University.

\end{document}